# Environmental Dependence of the Performance of Resistive Plate Chambers

Burak Bilki[d], John Butler[b], Ed May[a], Georgios Mavromanolakis[c,1], Edwin Norbeck[d], José Repond[a], David Underwood[a], Lei Xia[a], Qingmin Zhang[a,2]

[a]*Argonne National Laboratory, 9700 S. Cass Avenue, Argonne, IL 60439, U.S.A.*
[b]*Boston University, 590 Commonwealth Avenue, Boston, MA 02215, U.S.A.*
[c]*Fermilab, P.O. Box 500, Batavia, IL 60510-0500, U.S.A.*
[d]*University of Iowa, Iowa City, IA 52242-1479, U.S.A.*

**Abstract.** This paper reports on the performance of Resistive Plate Chambers (RPCs) as function of the gas flow rate through the chambers and of environmental conditions, such as atmospheric pressure, ambient temperature and air humidity. The chambers are read out by pads with an area of 1 x 1 $cm^2$ and a 1-bit resolution per pad. The performance measures include the noise rate as well as the detection efficiency and pad multiplicity for cosmic rays. The measurements extended over a period of almost one year and are sensitive to possible long-term aging effects.

**Keywords:** Calorimetry, Linear Collider, Particle Flow Algorithms, Resistive Plate Chambers. Environmental dependence, Gas flow rate

## INTRODUCTION

Particle Flow Algorithms (PFAs) [1], which attempt to measure each particle in a hadronic jet individually with the detector component providing the best resolution, require calorimeters with a finely segmented readout [2]. In this context, a small prototype hadron calorimeter with Resistive Plate Chambers (RPCs) as active elements and readout pads with an area of 1 x 1 $cm^2$, has been built and underwent extensive tests in particle beams [3-6].

This paper reports on long-term cosmic ray tests of the chambers utilized in the prototype calorimeter. During the period of twelve months, the noise rate (denoted by N), the detection efficiency (ε) and the pad multiplicity (μ) for cosmic rays were measured almost continuously. These performance variables were observed to fluctuate, sometimes even within short time intervals. In order to understand these fluctuations, correlations between the measurements and the environmental conditions, i.e. the atmospheric pressure (denoted by p), the ambient temperature (denoted by T) and air humidity (denoted by H), were investigated. In addition, the performance of the chambers was investigated as a function of the gas flow rate.

---

[1] Also affiliated with University of Cambridge, Cavendish Laboratory, Cambridge CB3 OHE, U.K.
[2] Also affiliated with Institute of High Energy Physics, Chinese Academy of Sciences, Beijing 100049, China and Graduate University of the Chinese Academy of Sciences, Beijing 100049, China.

Previously, several other groups studied the efficiency of RPCs as a function of temperature T and/or atmospheric pressure p [7-11]. Typically, these groups corrected the efficiency of the chambers by adjusting the applied high voltage V using a linear dependence: $V=V_0(T/T_0)(p_0/p)$ or $V=V_0(T/T_0)$, where $V_0$ is the nominal high voltage setting at a default temperature $T_0$ and atmospheric pressure $p_0$. One paper [10] parameterized the dependence of the efficiency on the atmospheric pressure as $\varepsilon = \varepsilon_0(1+\alpha\Delta p)$, where $\varepsilon_0$ is the efficiency at pressure $p_0$, $\alpha$ is a constant, and $\Delta p = p - p_0$. Another group [11] quotes the voltage at a 50% chamber efficiency as $V_{50\%}= V(1+\alpha\Delta T)(1-\beta\Delta p)$, where $\Delta T = T - T_0$ and $\alpha$, $\beta$ are constants.

This research was performed within the framework of the CALICE collaboration [12], which develops imaging calorimetry for the application of PFAs to the measurement of hadronic jets at a future lepton collider.

# THEORETICAL BACKGROUND

The response of an RPC to a traversing minimum ionizing particle depends on the number of initial ionizations in the gas gap and the ensuing electron multiplication or gas gain G. Due to the relative gas gain, even when operated in avalanche mode, the location of the first primary ionization in the gas gap is responsible for the broad signal charge distribution, typically from 100 fC to several pC. In the following we show that both the average location of the first primary ionization, under the assumption that the particle enters the gas gap through the cathode, and the gas gain depend on the density $\rho$ of the gas.

The number of primary ionizations dN in a slice dx of the gas gap is proportional to the density $\rho$ and the number of incident particles N not having initiated an ionization process

$$dN = -\alpha\rho N dx, \qquad (1)$$

where $\alpha$ is a proportionality constant. The average distance <d> of the first primary ionization to the anode plate can be calculated as

$$<d> = L - \frac{\int_0^\infty \frac{dN}{dx} x dx}{N_0} = L - \frac{1}{\alpha\rho} \qquad (2)$$

where $N_0$ is the number of incident particles and L is the thickness of the gas gap. As an approximation, the integral in Eq. (2) extends to infinity rather than to L only. Due to the high probability of having an ionization in the gas gap, this has only a negligible effect on the result.

According to J.Va'vra [13] the gain of chambers operated under atmospheric pressure varies depending on the density of the gas

$$\frac{dG}{G} = -\beta \frac{d\rho}{\rho}, \qquad (3)$$

where β is a proportionality constant with typical values between 5 - 8. After integration Eq. (3) becomes

$$G = \frac{G_0 \rho_0^\beta}{\rho^\beta} \propto \frac{1}{\rho^\beta}, \qquad (4)$$

where $G_0$ is the gain at a given gas density $\rho_0$.

Finally, it is relatively straightforward to show that if the <Q> signal charge or the charge distribution depends on the density of the gas, then so do the chamber's performance measures f: the efficiency ε, the average pad multiplicity μ and the noise rate N.

If the chamber's gas behaves similarly to an ideal gas, the density will be proportional to p/T, where p is the pressure and T the temperature[3]. Based on Eqs. (2) and (4) we therefore expect the performance variables $f_i$ (i= ε, μ or N) to depend only on the ratio p/T and to increase (decrease) with increasing temperature T (pressure p).

Given the moderate changes in temperature and atmospheric pressure during the tests, f(T/p) can be approximated using only the first term of the expansion in powers of T/p:

$$f(\frac{T}{p}) \approx f(\frac{T_0}{p_0}) + f'(\frac{T_0}{p_0}) \times (\frac{T_0}{p_0} \times \frac{\Delta T}{T_0} - \frac{T_0}{p_0} \times \frac{\Delta p}{p_0}) = f(\frac{T_0}{p_0}) + b_T \Delta T + b_P \Delta p \qquad (5)$$

$$\text{where } b_T = f'(\frac{T_0}{p_0}) \times \frac{T_0}{p_0} \times \frac{1}{T_0}, b_P = -f'(\frac{T_0}{p_0}) \times \frac{T_0}{p_0} \times \frac{1}{p_0}$$

$$\text{and } f = N, \varepsilon, \mu.$$

Here the standard pressure $p_0$ (temperature $T_0$) is taken to be 100 kPa (22.5 $^0$C) and Δp and ΔT indicate the difference from these values. The ratio of the slope parameters $b_T$ and $b_p$ can then be calculated as the ratio of the standard pressure $p_0$ and temperature $T_0$:

---

[3] In our set-up the gas pressure and temperature inside the chambers are close to the air pressure and ambient temperature in the laboratory.

$$\left|\frac{b_T}{b_P}\right| = \frac{p_0}{T_0} = \frac{100 kPa}{295.65 K} = 338\, pa/K \tag{6}$$

Since the gas volume is sealed from the surrounding air and glass is non-porous, the performance of the RPCs is not expected to depend on the humidity of the surrounding air. Again, the dependence on changes in humidity are assumed to be proportional to $\Delta H = H - H_0$, with $H_0$ equal to 40%.

## DESCRIPTION OF THE SET-UP

The tests included eight RPCs each with an area of 20 x 20 cm$^2$. Seven of these chambers contained two glass plates with a thickness of 1.1 mm. One chamber contained only one glass plate, also with a thickness of 1.1 mm. In both cases the gas gap was maintained with fishing lines and insulating tubing with a diameter of 1.2 mm. Schematics of the two chamber designs are shown in Fig. 1 and Fig. 2, respectively.

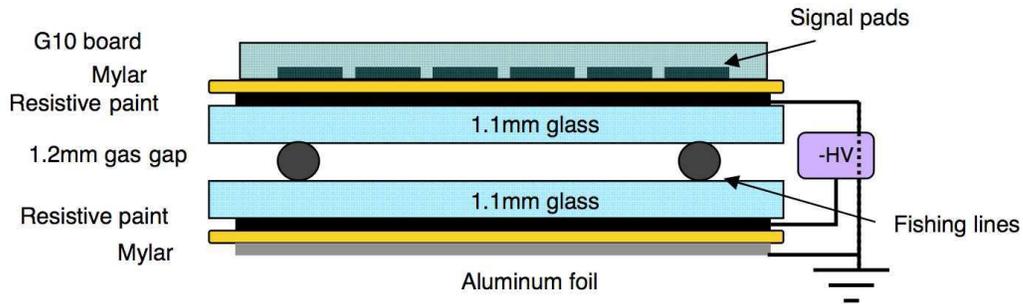

**Figure 1** Schematics of the 2-glass design RPC (not to scale).

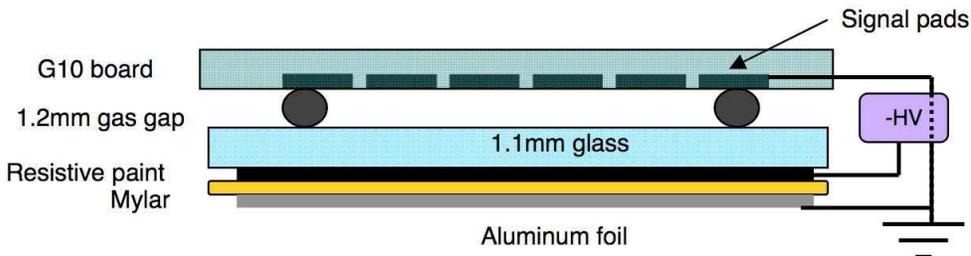

**Figure 2** Schematics of the 1-glass design RPC (not to scale).

The chambers were operated in saturated avalanche mode with a high voltage of 6.2 kV (6.0 kV) for the 2- (1-) glass chambers. The gas consisted of a mixture of three components: R134A (94.5%), isobutane (5.0%) and sulfur-hexafluoride (0.5%) [14].

The basic parameters of the eight chambers are summarized in Table 1. RPC1 and RPC2 were accidentally damaged by having the high voltage applied before being entirely flushed with the gas mixture. This accident resulted in significantly higher noise rates.

| RPC number  | 0    | 1       | 2       | 3    | 4    | 5    | 6    | 7    |
|-------------|------|---------|---------|------|------|------|------|------|
| Glass plates| 2    | 2       | 2       | 2    | 2    | 2    | 1    | 2    |
| HV [kV]     | 6.2  | 6.2     | 6.2     | 6.2  | 6.2  | 6.2  | 6.0  | 6.2  |
| Condition   | good | damaged | damaged | good | good | good | good | good |

**Table 1.** Basic information about the eight chambers participating in the tests.

The electronic readout system was optimized for the readout of large numbers of channels. Every chamber was read out by an array of 256 1 x 1 $cm^2$ readout pads. Each readout board contained four DCAL II chips [15], each of which served 64 readout pads. In order to avoid an unnecessary complexity of the electronic readout system, the charge resolution of individual pads was reduced to a single bit (digital readout). The threshold for registering a hit could be set in the range between 5 fC and 700 fC and was common to all channels of a given chip. Event data consisted of a time stamp (with a resolution of 100 ns) and a hit pattern. For more details on the readout system see [16].

The RPCs were mounted on plastic boards and stacked horizontally on top of each other. One scintillation counter on top and one underneath provided the triggers for cosmic rays. The trigger area was approximately 20 x 20 $cm^2$. Lead bricks with a thickness of 5 cm and located between the chambers and the bottom counter served as hardener for cosmic rays.

The environmental conditions were recorded using a weather station manufactured by Oregon Scientific [17]. Every 20 minutes the station recorded the time, the atmospheric pressure, the temperature (close to the stack), and the air humidity.

# DATA COLLECTION

Data was acquired in cycles of 120 minutes. Each cycle consisted of a 10 minute noise run and a 107 minute cosmic ray run. The remainder of the cycle was used for re-configuration and data backup.

## Noise runs

The measurement of the noise rate utilized the self-triggered data acquisition mode of the DCAL II chip [15]. All recorded hits in a run were summed up to yield the accidental noise rate of a given RPC. The statistical error of a 10-minute noise run was typically ~0.002 Hz/$cm^2$ or ~0.7% of the noise rate. In order to increase the sensitivity to the noise rate in the chambers, the data were taken at a relatively low threshold corresponding to approximately 60 fC. The expected hit rate from cosmic rays was estimated to be at least an order of magnitude smaller than the observed accidental rates.

Figure 3 shows the noise rate, the temperature and the atmospheric pressure as a function of time for selected chambers. It can be seen that in periods of relatively stable pressure the noise rates increased with increasing temperature. Vice versa, in periods of relatively stable temperature an indication that the rates decreased with increasing atmospheric pressure is visible.

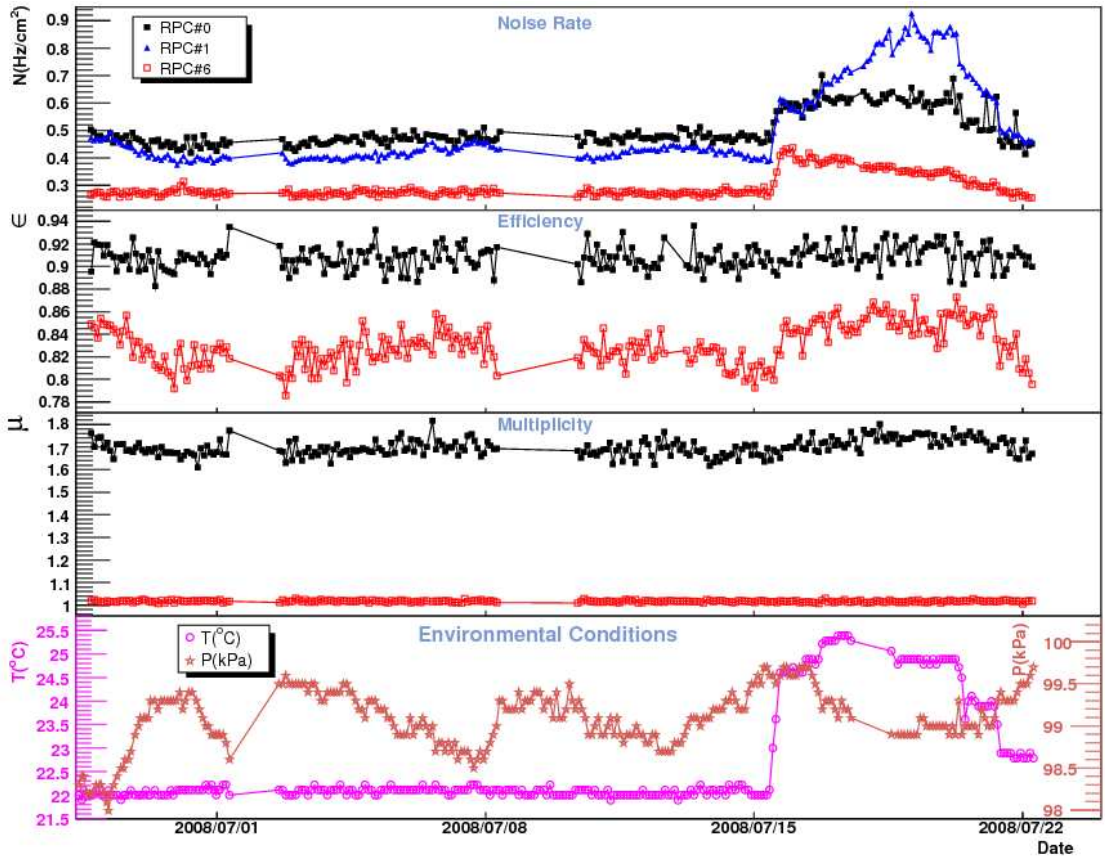

**Figure 3** Noise rate, efficiency, and multiplicity for selected chambers as a function of time. The lowest plot shows the variation of the atmospheric pressure and temperature during the same period of time. The efficiency and pad multiplicity of RPC1 is similar to RPC0 and is, therefore, not shown separately.

## Cosmic ray runs

Cosmic ray data were collected using the external trigger mode of the DCAL II chip. In a given event and in each layer of the stack, clusters were reconstructed as aggregates of cells with at least one side in common. The center of a cluster was defined as the arithmetic mean of the x and y coordinates of the pads belonging to that cluster. In a given cosmic ray event, each combination of seven (out of eight) RPCs was used to reconstruct a track through the centers of these clusters. These tracks were then interpolated or extrapolated to the one chamber not used in the track reconstruction, thus providing tracks for an unbiased measurement of the chamber's efficiency and pad multiplicity.

The detection efficiency of a given chamber was defined as the ratio of the number of events with at least one hit within 2.0 cm from the position of the reconstructed track to the total number of tracks. The pad multiplicity was calculated as the average number of pads in events with at least one hit within 2.0 cm from the position of the reconstructed track. In order to reduce a possible bias from multiple tracks through the chambers, the pad multiplicity was measured only for events where there was at most one cluster in that chamber. More details about this calculation can be found in [3].

The detection efficiency (pad multiplicity) for cosmic ray events as a function of time is also shown in Fig. 3. The statistical error on each data point is approximately 0.8% and 0.025 for the efficiency and pad multiplicity, respectively.

Whereas the efficiency of RPC0 (2-glass design) is seen to be quite stable, the efficiency of RPC6 (1-glass design) fluctuates with the environmental conditions. The difference in behavior of the two chambers is mostly due to the fact that RPC0 is operated on the efficiency plateau, whereas RPC6 is operated significantly below the plateau, thus increasing this chambers sensitivity to the environmental conditions.

Figure 3 also shows that the pad multiplicity of RPC6 is constant and close to unity, independent of the environmental conditions. On the other hand, the pad multiplicity of the two-glass chambers is seen to vary with the environmental conditions.

## DATA ANALYSIS AND RESULTS

In order to establish the correlation between the RPC performance variables and the environmental conditions, the measurements were corrected to a standard condition defined as $T_0 = 22.5\ ^0C$, $p_0 = 100$ kPa, and humidity $H_0 = 40\%$ using the linear approach outlined above. The correction procedure minimized the width of the $f_i$ distributions and established values for the slope parameters of the temperature and pressure dependences, $b_T$ and $b_p$.

### RMS Comparison

As a performance measure of the correction procedure the widths of the $f_i$ distributions were compared before and after the correction procedure. In the case of a perfect correction procedure, the widths of the $f_i$ distributions are expected to approach the width expected from the statistical fluctuations of the data. The following analysis treats the changes in humidity separately, since its effect is expected to be small.

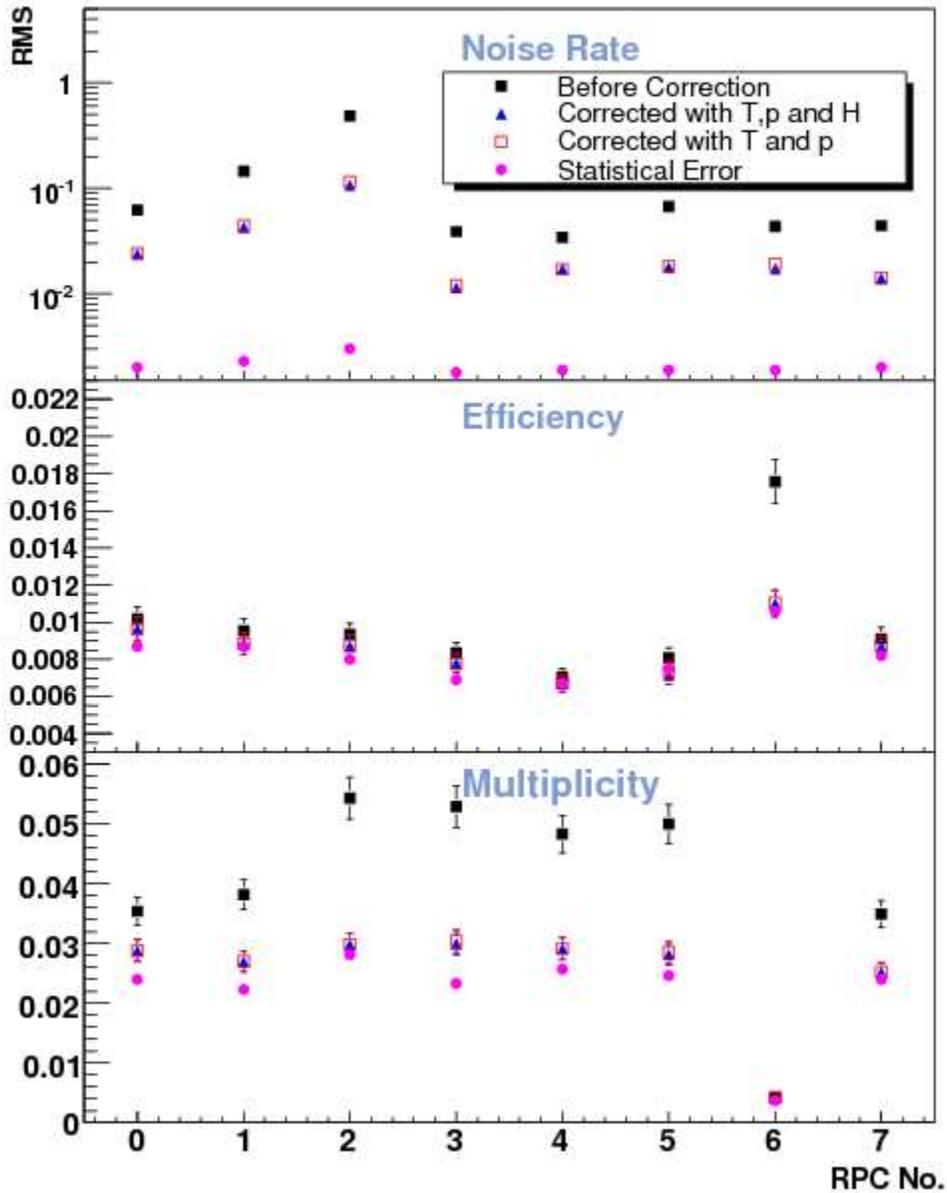

**Figure 4** RMS of the noise rate, detection efficiency, and pad multiplicity (from top to bottom) distributions before and after corrections for changes in temperature, pressure and humidity. The statistical errors are shown as magenta dots.

Figure 4 compares the widths of the $f_i$ distributions before corrections, after the $\Delta p$ and $\Delta T$ corrections, and finally after all three corrections, including the $\Delta H$ correction.

The correction for $\Delta H$ is seen to have a minimal effect on the widths of the distributions. This finding was expected as the gas system is sealed from the surrounding air and the glass, unlike Bakelite (the commonly used alternative to glass in RPCs), is not affected by humidity. On the other hand, the $\Delta p$ and $\Delta T$ corrections are seen to reduce the width of the distributions significantly.

## Slope parameters

Figure 5 shows the noise rate, detection efficiency, and pad multiplicity vs. ΔT and Δp for three selected RPCs. The results plotted against ΔT (Δp) have been corrected for the pressure (temperature) dependence. A positive (negative) slope of the temperature (pressure) dependence is clearly visible.

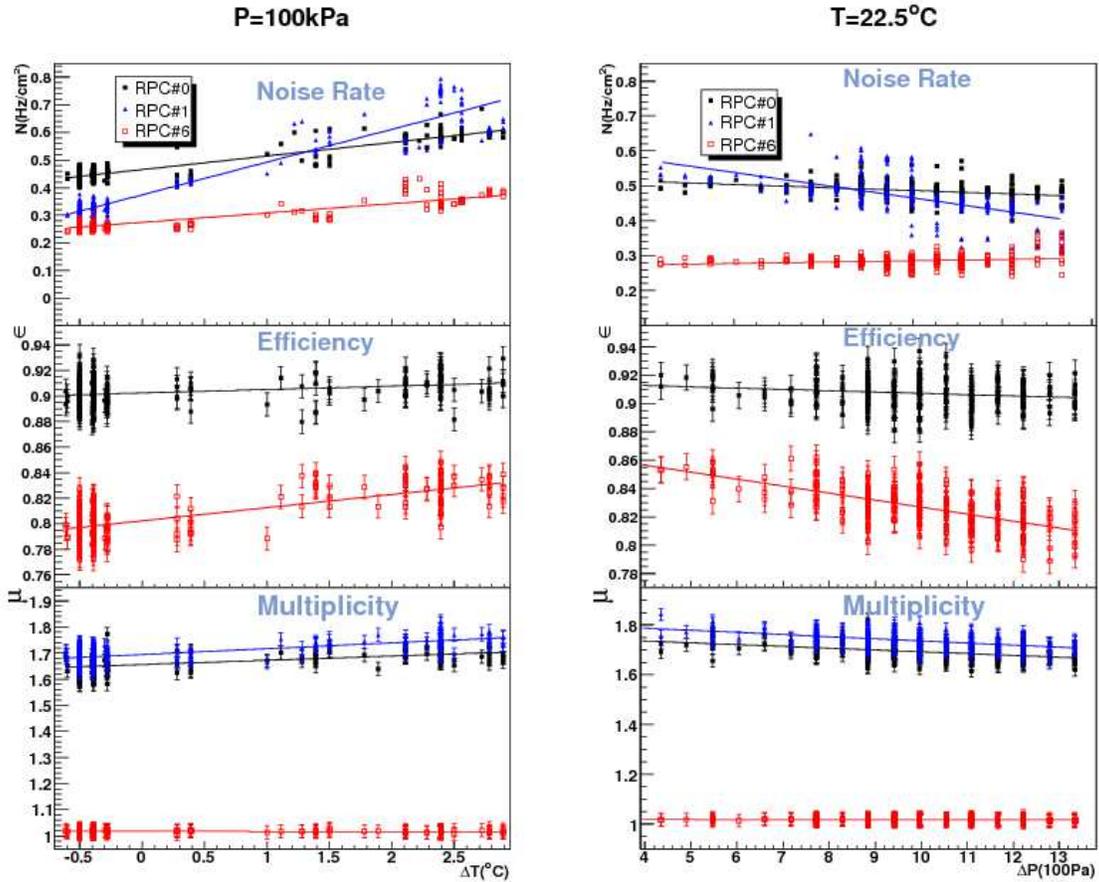

**Figure 5** Noise rate, detection efficiency, and pad multiplicity vs. ΔT (left) and Δp (right) for selected RPCs. In the right (left) plot the values have been corrected for the pressure (temperature) dependence. Since the results for the efficiency of RPC0 and RPC1 overlap, only the results for RPC0 are shown. The lines indicate the fitted slopes of the temperature and pressure dependences.

The resulting slopes, $b_T$ and $-b_P$, correspond to ΔT (Δp) equal to 1 $^0$C (-100 Pa) and are shown in Fig.6. Except for the noise rate of the damaged RPCs and the efficiency of the 1-glass RPC, similar $b_T$ and $b_P$ values are obtained among the various RPCs. Table 2 summarizes the average temperature and pressure dependences for the 1- and 2-glass chambers. The dependencies of the pad multiplicities of the 1-glass chamber are consistent with zero.

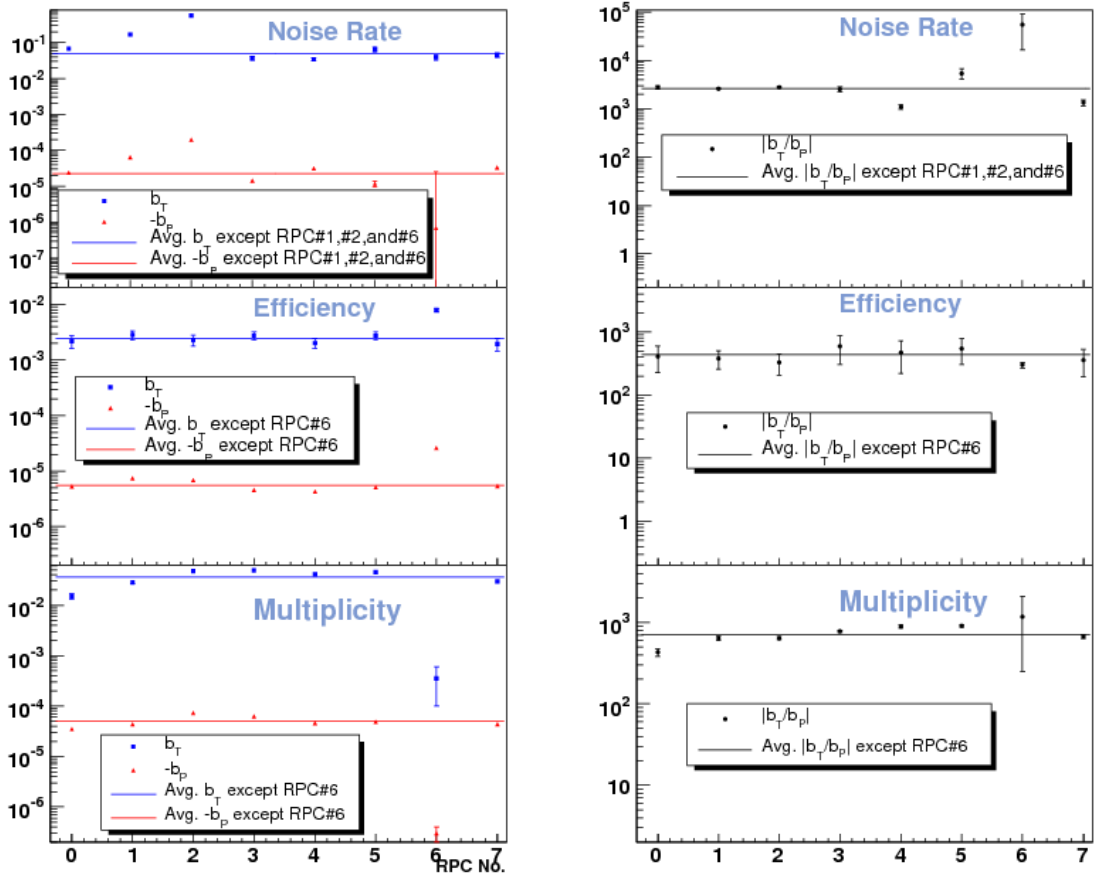

**Figure 6.** Left: parameters $b_T$ and $-b_P$ as determined from fits to the dependence of the noise rate, efficiency and multiplicity on temperature and pressure versus RPC number. Right: ratio of $b_T$ and $-b_P$ versus RPC number.

| Performance variable | Changes (in %) for $\Delta T = 1\ ^0C$ | | | Changes (in %) for $\Delta p = -100$ Pa | | |
|---|---|---|---|---|---|---|
| RPC design | 2-glass | | 1-glass Good | 2-glass | | 1-glass Good |
| | Good | Damaged | | Good | Damaged | |
| Noise rate | 14±1.6 | 42±1.2 | 13±1.8 | 0.70±0.037 | 1.73±0.028 | 0.02±0.694 |
| Efficiency | 0.26±0.051 | 0.28±0.0559 | 0.98±0.078 | 0.06±0.001 | 0.08±0.001 | 0.32±0.001 |
| Pad multiplicity | 2.0±0.09 | 2.0±0.09 | 0.035±0.0250 | 0.30±0.002 | 0.26±0.002 | 0.003±0.0010 |

**Table 2.** Relative change in performance for $\Delta T$ ($\Delta p$) = 1 $^0C$ (-100 Pa).

For the detection efficiency, the ratio compares well with the theoretical value derived above. However, for the pad multiplicity (noise rate) the measured ratio is approximately a factor of two (seven) larger than predicted.

## Corrected data points

Figure 7 compares the original data points and the data corrected for the temperature and pressure dependence as a function of time. The correction is seen to have smoothed out the data and reduced the various bumps and dips in the measurements. However, in the case of the damaged RPCs, the noise rate is seen to have been overcorrected. This phenomenon is currently not fully understood and requires further study. As expected, the correction works well for the efficiency data of the 1-glass RPC, but has a minor effect on the efficiency of the 2-glass RPCs, which are operated on the plateau of the efficiency curve. The multiplicity corrections work well for the 2-glass RPCs, but have a negligible effect on the 1-glass RPC.

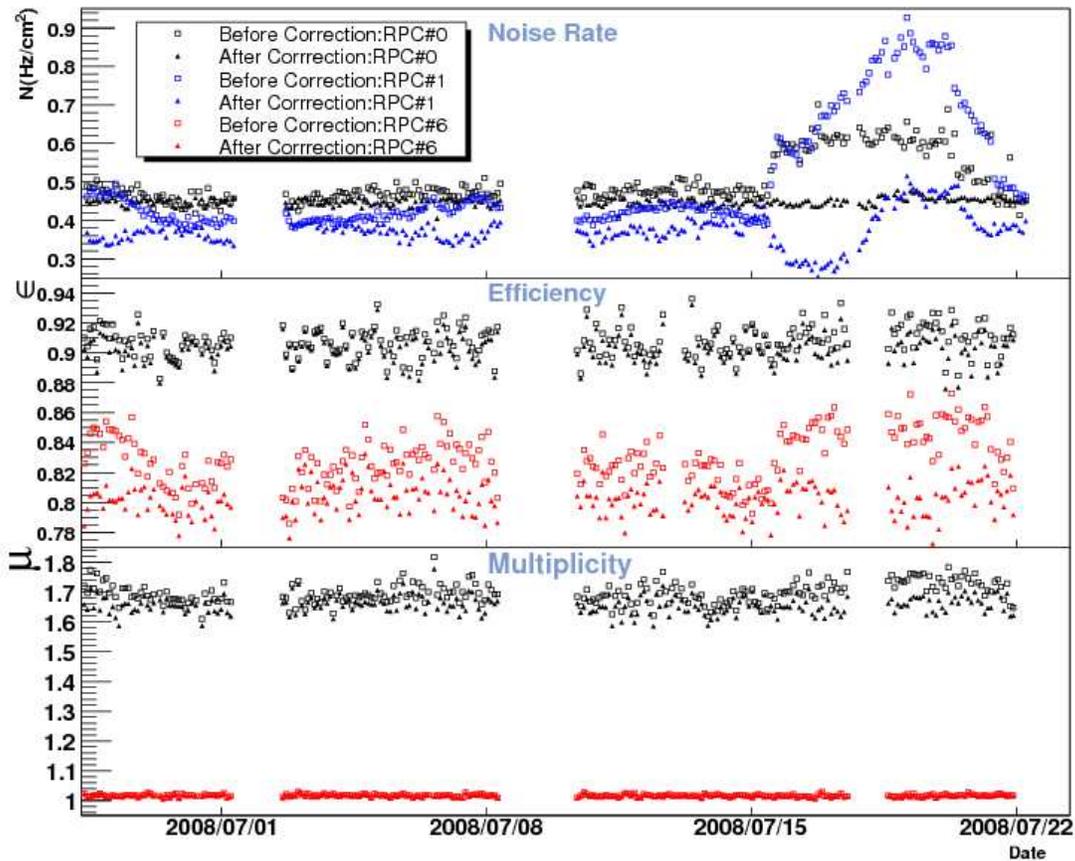

**Figure 7** Noise rates, efficiency and pad multiplicity of selected RPCs as a function of time before and after the corrections for the temperature and pressure dependences. For display purposes the noise rate of RPC6, the efficiency of RPC 1 and the multiplicity of RPC1 are not shown.

## Gas Flow Rate Test

Contamination of the gas may lead to a deterioration of the performance of the chambers. By continuously flushing the gas, contaminations produced through the ionization of gas molecules in avalanches can be removed.

The following study included RPC0 and RPC4 and investigated the effects of (extremely) low gas flow rates. The chambers were flushed with flow rates between 0.04 and 2.0 cc/min/RPC or between 1.2 and 60 volume changes per day. Three days after switching to a new flow rate, the noise rate, the efficiency and the pad multiplicity were measured. Figure 8 shows the performance variables as a function of gas flow rate. It is seen that below a rate corresponding to 8 volume changes per day, the noise rate and pad multiplicity increase rapidly with decreasing gas flow rate. In these tests the detection efficiency does no appear to be affected by the gas flow rate.

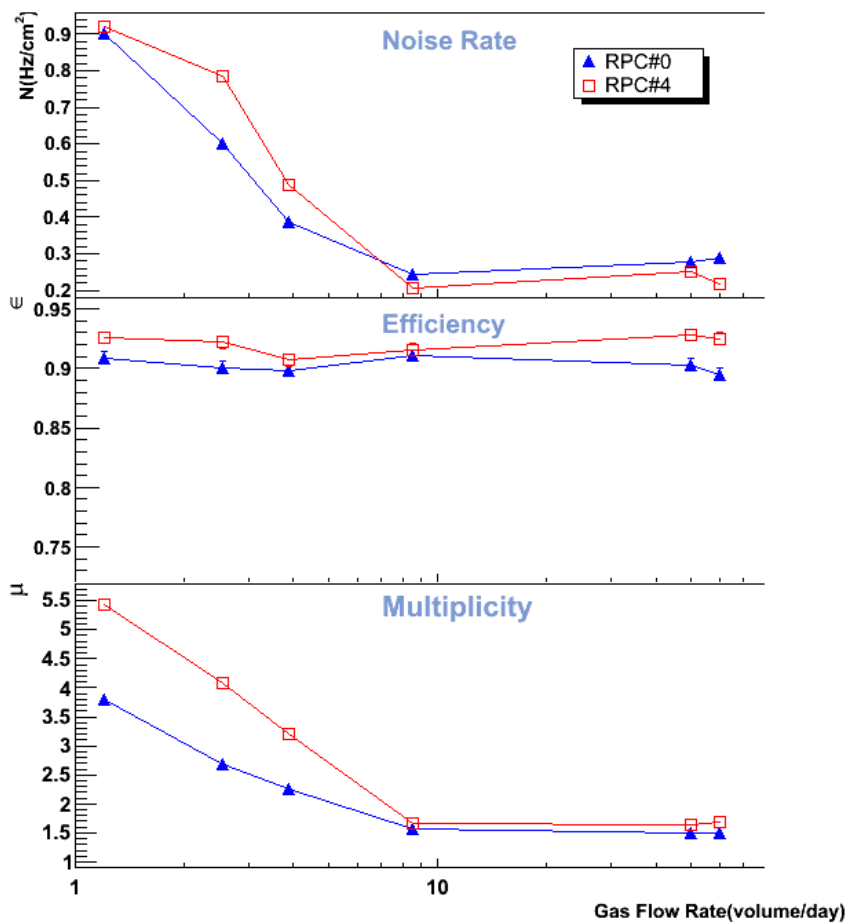

**Figure 8.** Noise rate, detection efficiency and pad multiplicity vs. gas flow rate.

Furthermore, it is seen that the performance for rates above about 8 volume changes per day, yields no further improvements.

## Long-term stability

The performance of the stack of RPCs was monitored during the period of one year to search for possible aging effects. Figure 9 shows the noise rate, detection efficiency and pad multiplicity for two 2-glass RPCs versus time. The periods in which there are no measurements coincide with safety reviews and periods of downtime for re-stacking and re-ordering of the chambers. The large increase in noise rate and pad multiplicity seen in December 2008 is related to the above mentioned studies with various gas flow rates. It is observed that the noise rate and pad multiplicity fell back to their usual values when the default flow rate was reinstated. The difference in pad multiplicity before and after January 2009 is related to different readout thresholds. Before this date the threshold was set to 80 DAC counts, corresponding to ~160 fC, and after that date the threshold was increased to 110 DAC counts, corresponding to ~220 fC. The efficiency appears to be insensitive to this change in threshold.

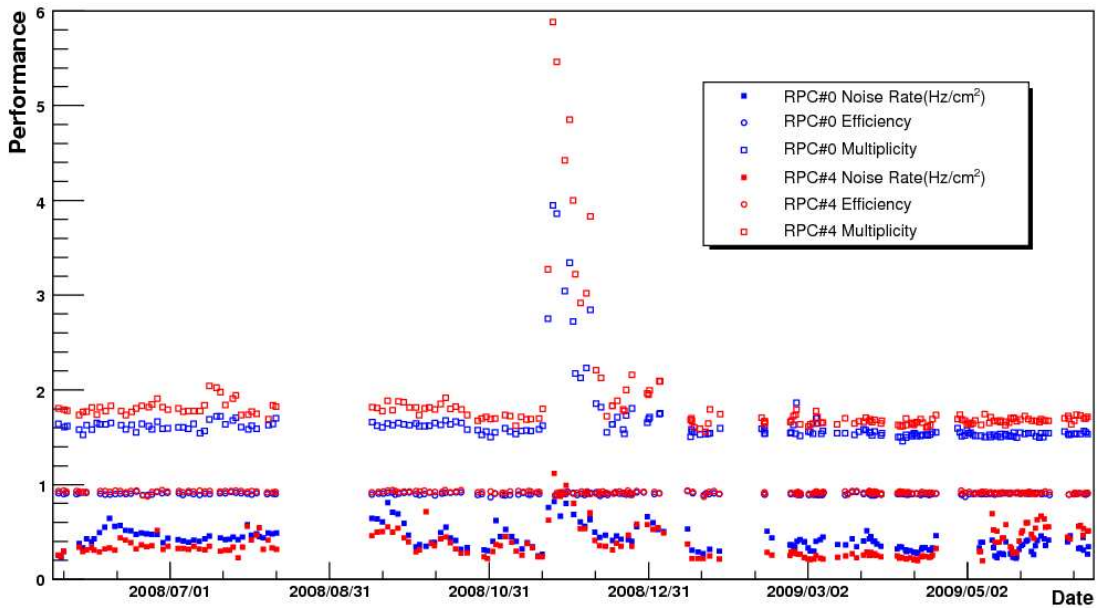

**Figure 9** Noise rate, MIP detection efficiency and pad multiplicity as a function of time for two 2-glass RPCs. Periods without measurements are due to safety reviews and re-stacking of the chambers. The large increases in noise rate and pad multiplicity seen in December 2008 are due to tests with various gas flow rates. The difference in multiplicity before and after January 2009 is a result of different readout thresholds during these periods of time.

## CONCLUSIONS AND DISCUSSIONS

The performance (noise rate, MIP detection efficiency and pad multiplicity) of Resistive Plate Chambers (RPCs) was measured as a function of environmental conditions (atmospheric pressure, ambient temperature and air humidity). The performance variables were found to be insensitive to the air humidity.

*Noise rate:* The measurements show a strong dependence of the noise rate on temperature and pressure and the corrections for the environmental conditions are seen to work well. However, there is an overcorrection phenomenon for damaged RPCs, which is currently not fully understood.

Changes in the rate of accidental hits can have two main causes: changes of the gas gain and contamination of the gas. For the good RPCs, it is assumed that the latter plays only a minor role.

The large value for the ratio of $b_T$ and $b_p$ suggests that indeed another effect, besides changes in the gas gain, may effect the noise rate. A possible temperature dependent mechanism might involve field emission of the cathode, thus creating seeds for accidental avalanches. This explanation is somewhat supported by independent studies of the temperature dependence of electrons escaping from the surface of materials [18, 19].

*Efficiency:* The 2-glass RPCs were operated on the plateau of the efficiency curve and are therefore less sensitive to variations in the gas gain. Accordingly, the dependence of the efficiency on the environmental conditions is observed to be small.

The 1-glass RPC was operated below the plateau of the efficiency curve and is consequently more sensitive to changes in the gas gain. The measured variations in efficiency have been studied in detail and can be explained through changes in the gas gain.

*Pad multiplicity:* For the 2-glass RPCs, the atmospheric pressure and temperature are the dominant factors affecting the pad multiplicity. The fluctuations, however, can not be explained solely as due to changes in the gas gain. Further studies are necessary to identify the causes of the larger than expected temperature dependence.

The 1-glass RPC shows a constant pad multiplicity close to unity and independent of environmental conditions.

*Gas flow rate:* Studies of the performance as function of gas flow rate showed no effect on the MIP detection efficiency, but a dramatic increase in noise rate and pad multiplicity for flow rates below 8 volume changes per day. This effect is most likely related to the contamination of the gas through ionization of the gas molecules by avalanches. The exact mechanism is, however, not yet fully understood and is still under study.

*Long term effects:* In twelve months of almost continuous operation of a stack of RPCs, no degradation of the performance has been observed.

# ACKNOWLEDGEMENT

We would like to thank Crispin Williams from the INFN Bologna for useful discussions.

# REFERENCES

1. M.A. Thomson, Particle Flow Calorimetry and the PandoraPFA Algorithm, arXiv: 0907.3577, submitted to Nucl. Instrum. Meth.
2. J. Repond, Proceedings of the 10th Pisa Meeting on Advanced Detectors, La Biodola, Isola d'Elba, Italy, Nucl. Instrum. Meth, A572, 211 (2007).
3. B. Bilki et al., Calibration of a Digital Hadron Calorimeter with Muons, 2008 JINST 3 P05001.
4. B. Bilki et al., Measurement of Positron Showers with a Digital Hadron Calorimeter, 2009 JINST 4 P04006.
5. B. Bilki et al., Measurement of the Rate Capability of Resistive Plate Chambers, 2009 JINST 4 P06003.
6. B. Bilki et al., Hadron Showers in a Digital Hadron Calorimeter, 2009 JINST 4 P10008.
7. G. Aielli et al, Large Scale Test and Performances of the RPC Trigger Chambers for the Atlas Experiment at the LHC, Nuclear Science Symposium Conference Record, 2004 IEEE Vol. 1, P538- 542.
8. K. Doroud et al, Simulation of temperature dependence of RPC operation, Nucl. Instrum. Meth, A602 723–726 (2009).
9. M. De Vincenzi et al., Study of the performance of standard RPC chambers as a function of Bakelite temperature, Nucl. Instrum. Meth, A508 (2003) 94–97.
10. R. De Asmundis et al, Using RPC Detectors as Cosmic Rays Monitors, IEEE Transactions on Nuclear Science, VOL. 54, NO. 3, June 207, P670-676.
11. M. Bianco et al., ATLAS RPC certification with cosmic rays, Nucl. Instrum. Meth, A602 700–704 (2009).
12. https://twiki.cern.ch/twiki/bin/view/CALICE/WebHome
13. J. Va'vra, Gaseous Wire Detectors, SLAC-PUB-7627 (August 1997).
14. P.Camarri et al., Streamer suppression with SF in RPC's operated in avalanche mode, Nucl. Instrum. Meth. A414, 317 (1998).
15. J. Hoff et al., A custom integrated circuit for calorimetry at the International Linear Collider, IEEE Nucl. Sci. Symp., Puerto Rico, 2005, FERMILAB-CONF-05-509.
16. J. Butler et al., A new readout system for digital hadron calorimetry for the International Linear Collider, IEEE Nucl. Sci. Symp. Conf. Rec. 3 (2007) 2145, submitted to IEEE Trans. Nucl. Sci.
17. http://www.oregonscientific.com
18. S. Chen et al, Temperature dependence of surface band bending and field emission for boron-doped diamond and diamond-like films, New Journal of Physics 4 (2002) 79.1-79.7.
19. A. Tataroğlu et al, Effect of Surface States on Electrical Characteristic of Metal-Insulator-Semiconductor, ISSN 1303-9709, 16(4):677-685, 2003.